\documentclass[aps, prl, floatfix, twocolumn, superscriptaddress]{revtex4-1}
\usepackage{calc}
\usepackage{lineno}
\usepackage{xcolor}
\usepackage{siunitx}
\usepackage{hyperref}
\usepackage{graphicx}
\usepackage{mathrsfs}
\usepackage{mathtools}
\usepackage{amsmath, amssymb, amsfonts, amsbsy, bm}
\usepackage{braket}
\usepackage{bbm}

\DeclarePairedDelimiter\abs{\lvert}{\rvert}%
\begin{document}

\title{Momentum-space signatures of the Anderson transition \\
       in a symplectic, two-dimensional, disordered ultracold gas}

\author{Ehsan  \surname{Arabahmadi}}
\email{ehsan.arabahmadi@postgrad.otago.ac.nz}

\author{Daniel \surname{Schumayer}}
\affiliation{Dodd-Walls Centre for Photonic and Quantum Technologies,
             Department of Physics, University of Otago, Dunedin, New Zealand}

\author{Beno{\^\i}t \surname{Gr{\'e}maud}}
\affiliation{Aix Marseille Universit{\'e}, Universit{\'e} de Toulon,
             Centre National de la Recherche Scientifique, CPT,
             Marseille, France}
\affiliation{MajuLab, CNRS-UCA-SU-NUS-NTU International Joint Research Unit,
             Singapore}
\affiliation{Centre for Quantum Technologies, National University of Singapore,
             Singapore}

\author{Christian \surname{Miniatura}}
\affiliation{MajuLab, CNRS-UCA-SU-NUS-NTU International Joint Research Unit,
             Singapore}
\affiliation{Centre for Quantum Technologies, National University of Singapore,
             Singapore}
\affiliation{Department of Physics, National University of Singapore, Singapore}
\affiliation{School of Physical and Mathematical Sciences,
             Nanyang Technological University, Singapore}
\affiliation{Universit{\'e} C{\^o}te d'Azur, CNRS,
             Institut de Physique de Nice, France}

\author{David A.~W. \surname{Hutchinson}} 
\affiliation{Dodd-Walls Centre for Photonic and Quantum Technologies,
             Department of Physics, University of Otago, Dunedin,
             New Zealand}
\affiliation{Centre for Quantum Technologies,
             National University of Singapore, Singapore}

\begin{abstract}
    We study Anderson Localization in two dimensional (2D) disordered spin-orbit systems described by the Gaussian symplectic ensemble using momentum-space signatures such as the coherent backscattering (CBS) anti-peak, and the coherent forward scattering (CFS) peak. Significantly, these momentum-space features are readily accessible in ultracold atom experiments through absorption imaging after time-of-flight expansion. The critical exponent and mobility edge of the metal-insulator transition are successfully obtained in this model through a finite-time analysis of the CBS width. An anomalous residual diffusion, unique to 2D, is identified at the transition point where the system changes from a metal to an insulator. A spin localization phenomenon is also observed in the deep localized regime.
\end{abstract}

\date{\today}
\maketitle

\emph{Introduction}---
    Anderson localization (AL), the disorder-induced suppression of wave transport by destructive interference, was first introduced \cite{Anderson1958} to explain the anomalous suppression of conductance in mesoscopic electron systems.
\begin{table*}
    \caption{\label{tab:SymmetriesAndDimensionality}
             Phases in symmetry classes and dimensions. Abbreviations: metal-insulator transition (MIT), only localized states (L). Corresponding review articles are Refs.~\cite{Slevin2014a, Ueoka2017, RevModPhys.80.1355}
            }
    \begin{ruledtabular}
    \begin{tabular}{lllllll}
         symmetry & $d=1$ & $d=2$ & $d=3$ & $d>3$ & system \\
         \hline
         orthogonal &
         L   &
         L   &
         MIT &
         MIT &
         no spin-orbit coupling, no magnetic field
         \\
         symplectic &
         L   &
         MIT &
         MIT &
         MIT &
         spin orbit coupling
         \\
         unitary &
         L   &
         L   &
         MIT   &
         MIT &
         magnetic field
         \\
    \end{tabular}
    \end{ruledtabular}
\end{table*}
It is, in fact, a general phenomenon, and an ubiquitous feature of any linear waves propagating in bulk random media. Since its conceptual inception, it has been observed (if indirectly) in a variety of very different systems \cite{Mott1969, Lee1985, Weaver1990, McCall1991, Wiersma1997, Genack1997, Weiland1999, Storzer2006, Topolancik2007, Schwartz2007, Hu2008, Chabe2008, Riboli2011, Sperling2012, Lopez2012, Manai2015, Ying2016}. Notably, over the past decade, ultracold atomic gases have provided a uniquely controllable experimental platform in which to directly observe and study AL in quantum systems \cite{PhysRevLett.109.195302, Roati2008, doi:10.1126/science.1209019, PhysRevLett.105.090601, PhysRevLett.101.255702, Billy2008, Semeghini2015}.

In particular, the momentum distribution of the single-particle wavefunction has provided a directly observable signature of both weak localisation, and strong localisation through the coherent back-scattering (CBS) and coherent forward-scattering (CFS) peaks \cite{Muller2010, Ghosh2014, Ghosh2015}. Their dynamic observation can be used to quantitatively characterise the three dimensional (3D) Anderson transition delineating an extended metallic regime from an insulating one \cite{Ghosh2014, Ghosh2015}.

Historically, the first powerful phenomenological description of AL was the one-parameter scaling theory \cite{Thouless1974, Abrahams1979}. It relies on the hypothesis that all transport properties of a disordered system depend only on the dimensionless conductance $g$. The scaling behavior of $g$ with the system size $L$ is encapsulated in the function $\beta(g) = \frac{d\ln g}{d\ln L}$, and obtained from a smooth interpolation between the limiting metallic and insulating expected asymptotics. This theory predicts the existence of a metal-insulator transition (MIT) in dimension 3 \cite{Mott1987, Lee1985}. It was also conjectured that there are distinct universality classes based on the symmetries of the Hamiltonian: orthogonal, unitary, and symplectic. For example, ultracold atoms spreading in an optical speckle potential, where both time and spatial rotational symmetries are present, are well described by the Gaussian Orthogonal Ensemble (GOE) of random matrix theory \cite{asada2005anderson}. It is also well-known that disordered systems within this symmetry class are always localised for any disorder strength in dimension two or less, whereas they exhibit a metal-insulator transition in dimension three. In particular, both the mobility edge and critical exponent of this Anderson transition have been determined through the scaling behaviors of the CBS width and CFS contrast \cite{Gosh2014a, Gosh2015a}. 

On the other hand, AL within the Gaussian Unitary (GUE) and Symplectic (GSE) Ensembles has received less experimental attention in the ultracold atom community \cite{Slevin2014a, Ueoka2017, RevModPhys.80.1355, Asada2002a}. In this Letter, we address the GSE case by considering spin-$\tfrac{1}{2}$ particles in a two dimensional (2D) square lattice with onsite disorder and random spin rotation during hopping. As is well known, spin-orbit (SO) coupling induces a MIT in two dimensions at low enough disorder \cite{RevModPhys.80.1355,Orso2017}. We use the scaling properties of the CBS (anti)-peak present in the momentum distribution of the particles to extract the mobility edge and critical exponent of this transition. The scaling behaviour of the CFS peak contrast will be addressed in future work.

\emph{Theoretical model}---
Our tight-binding Hamiltonian for noninteracting spin-$\tfrac{1}{2}$ particles reads:
\begin{equation}
    \label{eq:Hamil}
    H
    = J \, \sum_{\langle i, j \rangle}%
        \psi_{i}^{\dagger} {U_{ij} \psi_{j}^{\phantom{\dagger}}}
        + \sum_{i}{w_{i} \psi_{i}^{\dagger} \psi_{i}^{\phantom{\dagger}}},
\end{equation}
where the sums run over all nearest-neighbor lattice site pairs $\langle i, j \rangle$ and lattice sites $i$, respectively. The field operator $\psi_{i}^{\dagger} = (\psi^\dag_{i\uparrow},\psi^\dag_{i\downarrow})$ is the 2-component row-spinor built from the creation operators $\psi^\dag_{i\sigma}$ at site $i$ and spin components $\sigma = \,\uparrow$, and $\downarrow$. The onsite disorder potentials $w_{i}$ are independent random variables uniformly distributed over $[-W/2,W/2]$, where $W$ is the disorder strength. Hereafter, we set the hopping amplitude to $J=1$, the lattice spacing $a=1$ and $\hbar =1$. Following \cite{Asada2002a}, the random spin rotation during hopping is described by the $SU(2)$ matrix
\begin{equation}
    \label{eq:Hop}
    U_{ij}
    =
    \begin{bmatrix}
        \phantom{-}e^{\mathrm{i} \alpha_{ij}} \cos(\beta_{ij}) &
        e^{ \mathrm{i} \gamma_{ij}} \sin(\beta_{ij})
        \\
       -e^{-\mathrm{i} \gamma_{ij}} \sin(\beta_{ij}) &
        e^{-\mathrm{i} \alpha_{ij}} \cos(\beta_{ij})
    \end{bmatrix}
\end{equation}
where the angles $\alpha_{ij}$ and $\gamma_{ij}$ are independent random variables uniformly distributed over $[0, 2\pi)$ while the angles $\beta_{ij}$ are independent random variables distributed over $[0,\pi/2]$ with probability density function $g(\beta) = \sin(2\beta)$. Since $H$ is Hermitian, $U_{ij} = U^\dag_{ji}$ implying $\alpha_{ij} = -\alpha_{ji}$, and similarly for $\gamma_{ij}$ and $\beta_{ij}$. 
 
One recovers the GOE case for $\beta_{ij}=0$ and constant uniform angles $\alpha_{ij}$ and $\gamma_{ij}$. Noticeably, $H$ is invariant under time reversal, $T H T^{-1} =H$, where $T$ is the time reversal operator for spin-$\tfrac{1}{2}$ systems and satisfying $T^{2} = -\mathbbm{1}$ \footnote{For spin-$\tfrac{1}{2}$ system $T = -\mathrm{i} \sigma_y \, K$ , where $\sigma_y$ is the $y$-Pauli matrix and $K$ the complex conjugation operator. Therefore $T^{2} = -\mathbbm{1}$}. As a consequence, each eigenvalue $\varepsilon_{n}$ of $H$ is doubly degenerate (Kramers' degeneracy) with orthonormal eigenstates of the form $\ket{\varphi_{n}}$ and $\ket{T\varphi_{n}}$. 

Importantly, the Hamiltonian dynamics cannot couple time-reversed states, irrespectively of the disorder configuration. Indeed
\begin{align*}
    \bra{T \psi} & e^{-\mathrm{i} Ht} \ket{\psi}
    =
    \\
    &\sum_{n}
        {e^{-\mathrm{i} \varepsilon_{n} t}
         \Bigl \lbrack 
            \braket{T\psi | \varphi_{n}} \braket{\varphi_{n} | \psi}
            +
            \braket{T\psi| T \varphi_{n}} \braket{T \varphi_{n} | \psi}
        \Bigr \rbrack
       }.
\end{align*}
Using the $\braket{ T \psi | T \varphi_{n}} = \braket{\varphi_{n} | \psi}$ relationship together with $\braket{T\varphi_{n} | \psi} = - \braket{T \varphi_{n} | T^{2} \psi} = - \braket{T \psi | \varphi_{n}}$, we see that the bracketed term in the sum above vanishes. As will be seen later, this very fact explains why a CBS dip, rather than a CBS peak, is observed in the momentum distribution for GSE systems. 

\emph{Momentum distributions}---
To study the momentum-space signatures of AL, we consider the initial plane wave state $\ket{\boldsymbol{k}_{0}, \uparrow}$ at wave vector $\boldsymbol{k}_{0} = (0, \pi/2)$ that we shape into a wave packet $\ket{\psi_0} = \mathcal{F}(E, \delta E)\ket{\boldsymbol{k}_{0}, \uparrow}$ at energy $E$ by the filter operator $\mathcal{F}(E, \delta E) \propto \exp{\!\left [ -(H-E)^{2}/(2 \delta E^{2}) \right]}$. The parameter $\delta E$ that controls the selected energy window around $E$ should be as narrow as possible and simultaneously wide enough to keep a sufficient number of eigenstates \cite{Sanjib2017}. We then compute the disorder-averaged momentum distributions $n_{\sigma}(\boldsymbol{k},t) = \overline{ \abs{\braket{\boldsymbol{k}, \sigma | \exp(-\mathrm{i} H t) |\psi_0}}^{2}}$ at energy $E$ ($\sigma=\uparrow,\downarrow$). In the rest of the paper, we have chosen $E=1$ and $\delta E = 0.035$ (in units of $J$).

\begin{figure}[b]
    \includegraphics[width=82mm]{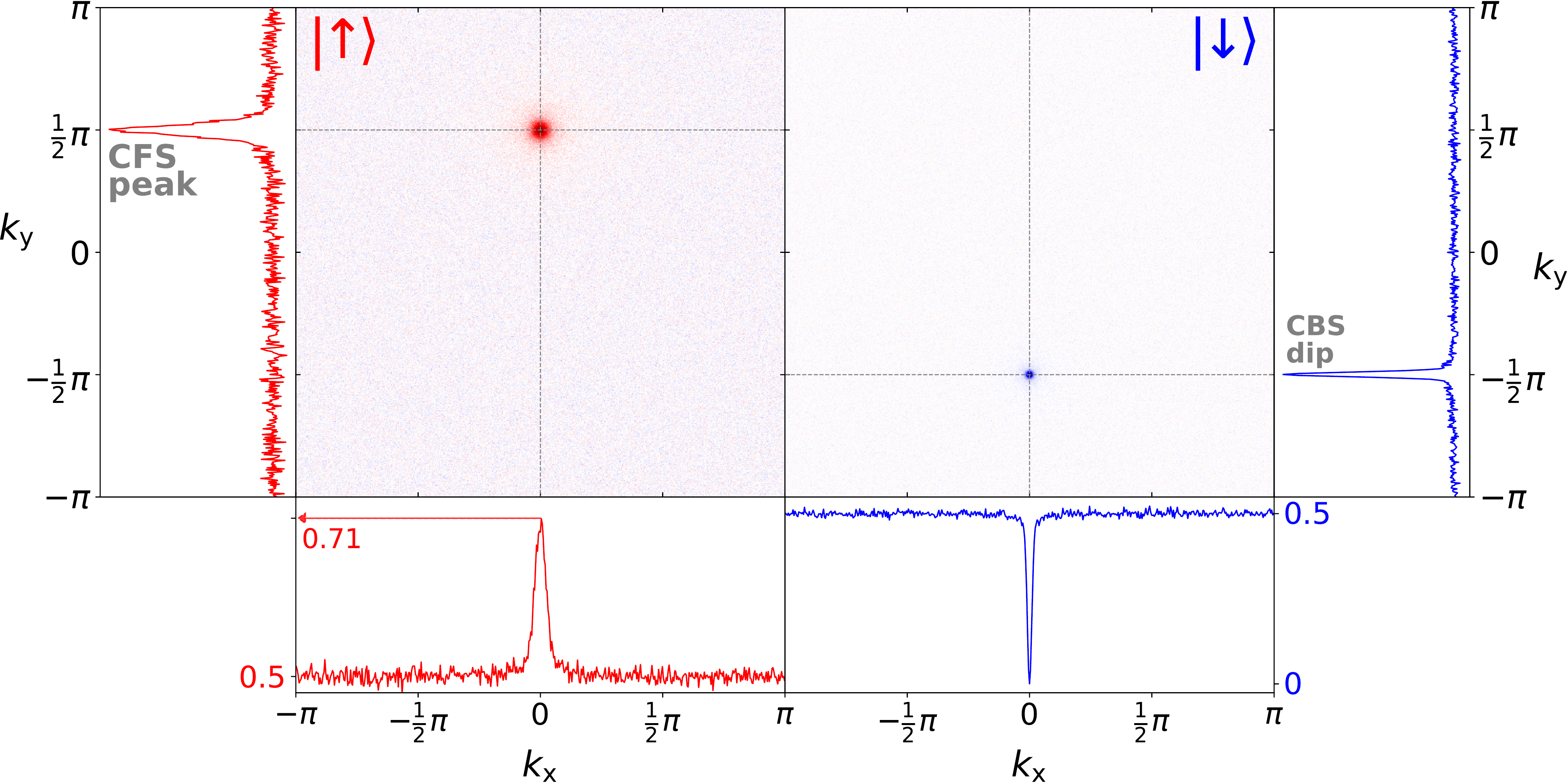}
    \caption{\label{fig:MomentumDistributions2D}
             Momentum distributions $n_{\uparrow}(\boldsymbol{k},t)$ and $n_{\downarrow}(\boldsymbol{k},t)$ obtained at time $t=100 \ \hbar/J$ for an initial state $\ket{\boldsymbol{k}_0,\uparrow}$ with $\boldsymbol{k}_{0} = (0, \pi/2)$  filtered at energy $E=1$ (in units of $J$). The linear size of the lattice is $L=513$ (in units of $a$) and the onsite disorder strength is $W=6.8$ (in units of $J$). The CFS peak and the CBS dip are clearly seen in their respective spin channels. At $t=100 \hbar/J$, the backgrounds in each spin channel have already reached their stationary and equal values (set to $1/2$ by total probability conservation). However, we note that the CFS contrast has not yet reached the stationary value $C^\infty_F=2$ expected for GSE systems.
           }
\end{figure}

Fig.~\ref{fig:MomentumDistributions2D} shows the momentum distributions obtained at time $t=100$ (in units of $\hbar/J$) for onsite disorder strength $W=6.8J$ (localized phase as will be seen later). In the spin-preserving channel, we observe a CFS peak centered at $\boldsymbol{k}_{0}$ on top of a flat diffusive background. In the spin-flipping channel, we observe a CBS anti-peak centered at $-\boldsymbol{k}_{0}$ and dug into a flat background. Since $\ket{-\boldsymbol{k}_{0},\downarrow} = T\ket{\boldsymbol{k}_{0},\uparrow}$, the dynamics cannot connect these two states and $n_\downarrow(-\boldsymbol{k}_{0},t) = 0$ at any time, irrespective of the disorder configuration averaging. The CBS dip is thus a genuine characteristics of GSE systems.

In addition, we note that both backgrounds in each spin channels are flat. This can be traced back to the fact that the disorder-averaged Green's function $\overline{G(E)} = \overline{(E-H)^{-1}}$, which is a diagonal operator in momentum and spin spaces as disorder average restores translation and rotation invariances, has diagonal elements that do not depend on ${\bm k}$ and $\sigma$ but only on $E$, i.e., $\langle \mathbf{k},\sigma|\overline{G(E)}|\mathbf{k},\sigma\rangle = \bar{g}(E)$. This unusual property, that we have checked numerically, can be explained by the fact that the disorder-averaged Hamiltonian vanishes ($\overline{H} = 0$), amounting to having a trivial ${\bm k}$-independent diagonal disorder-free Green's function $\langle {\bm k}'\sigma'|G_{0}(E)|{\bm k}\sigma\rangle =\delta_{{\bm k}{\bm k}'}\delta_{\sigma\sigma'}/(E+i0^+)$, and by the fact that the various correlators $\langle{\bm k}\sigma|\overline{H^{n}}|{\bm k}\sigma\rangle$, appearing in the Dyson series, are independent of ${\bm k}$ for the uncorrelated hopping and on-site independent disorders that we consider here. A proof, for Gaussian disorder, can be found in the Appendix of \cite{Martinez2022}. This has to be contrasted with the standard situation of onsite disorder only where the disorder-averaged Hamiltonian exhibits a well-defined band structure $\epsilon_{\mathbf{k}}$ in momentum space. This entails ${\bm k}$-dependent diagonal elements of the free Green's function $\langle {\bm k}\sigma|G_0(E)|{\bm k}\sigma\rangle = (E-\epsilon_{\mathbf{k}}+i0^+)^{-1}$.  Finally, from both  diagrammatic approach and numerical computations, one can show that $\bar{g}(E) = (E-\Sigma(E))^{-1}$, where the complex-valued scalar $\Sigma(E)$ is the self-energy. Therefore, one expects not only the backgrounds in each spin channel to be flat in the Brillouin zone, but also to grow with the same scattering time scale $\tau_s(E) = \hbar/(2|\textrm{Im}(\Sigma)|)$, before reaching the same stationary values.

Since $n_\downarrow(-\boldsymbol{k}_{0},t) = 0$ at any time, the flat diffusive background in the $\downarrow$-channel grows ``around'' the CBS dip. As time further increases, the CBS width shrinks and its temporal behaviour depends on whether the system is diffusive, localised, or critical. The CFS peaks develops and grows in the $\uparrow$-channel on a time-scale given by the localization time $\tau_{\text{loc}}$. It reaches a stationary peak-to-background relative contrast $C^\infty_{F}$ at ``infinite'' times, $t\gg \tau_{\text{loc}}$. Based on the statistical properties of the eigenfunctions in the GSE ensemble, we expect $C^\infty_{F}=2$ instead of the $C^\infty_{F}=1$ for GOE systems. Note that, in Fig.~\ref{fig:MomentumDistributions2D}, the momentum distributions are plotted at a time where the CFS peak has not yet reached its stationary value. Note also that deviations from the GSE value are expected when the localization length becomes too small, and comparable to the lattice constant, at large $W$ values~\cite{Ghosh2017,LKL2014}. On the other hand, by definition, the stationary CBS dip-to-background relative contrast is always $C^\infty_{B}=1$, like in the GOE case. In the remainder of this Letter, we will focus on the CBS dynamics and leave the discussion of the CFS dynamics to a forthcoming paper.

\emph{CBS width dynamics}---
We define the CBS width $\Delta k$ as the momentum size of the dip at half-maximum of the diffusive background in the spin-flipping channel. In the metallic regime, the CBS anti-peak continues to shrink in time and asymptotically tends to zero. At large enough times, its width is given by~\cite{Sanjib2017} 
\begin{align}
\label{eq:CBSDif}
    \Delta k(t)
    =
    \sqrt{\frac{\ln 2}{D(E,W)\,t}} 
    \quad \textrm{(metallic phase)},
\end{align}
where $D(E,W)$ is the diffusion constant at energy $E$ and disorder strength $W$. In the insulating regime, the CBS width decreases until it asymptotically approaches a constant value which defines the localisation length at energy $E$ and disorder strength $W$
\begin{align}
    \label{eq:CBSLoc}
    \Delta k (t \to +\infty)
    =
    \frac{1}{\xi_{\text{loc}}(E,W)}
    \quad \textrm{(insulating phase)}.
\end{align}

At fixed energy, $\xi_{\text{loc}} \sim |W-W_c|^{-\nu}$ diverges algebraically with a critical exponent $\nu$ when approaching the critical point $W_c$. Thus, $\xi_{\text{loc}}$ quickly exceeds the maximum linear size $L$ of the lattice that is computationally manageable and the system appears diffusive (in other words, $\Delta k$ sticks to the mesh size $2\pi/L$ in momentum space). This is the reason why we resorts to finite-time scaling methods \cite{Beltukov2017, Ghosh2014, Ghosh2015, Ghosh2017, PhysRevA.80.043626, Lemari2009} of $\Delta k$, and introduce the length scale $L_{t}$ through $t = 2\pi \rho(E) L^2_{t}$, where $\rho(E,W) = (1/L)^2 \, \overline{\sum_n \delta(E-\varepsilon_n)}$ is the disorder-averaged density of states (DoS) per unit surface of the system at energy $E$ and disorder strength $W$.

\emph{Finite-time scaling}---
Following the single-parameter scaling rationale \cite{Abrahams1979}, we assume that there exists a single correlation length $\xi$ subsuming all the microscopic details of the system. This correlation length identifies with the localization length in the insulating regime. As a consequence, the inverse of the rescaled CBS width is a continuous and smooth function of the single variable $L_{t}/\xi$ that we recast under the form: 
\begin{align}
    \label{eq:ScaledCBS}
    \Lambda \equiv [\Delta k \, L_t]^{-1} = F(z),
\end{align}
where $z = \eta(E,W) L_{t}^{1/\nu}$, $\eta(E,W) = \xi^{-1/\nu}$ and $F(z)$ is a function characteristic of the transition. Working at fixed energy, we now Taylor expand $F(z)$ and $\eta (E,W)$ up to some expansion orders, $F(z) = \sum_{n=0}^{N}{F_{n} z^{n}}$ and $\eta(E,W) = \sum_{m=1}^{M} {b_{m} (W-W_{c})^{m}}$ \cite{Ghosh2015,Asada2003a} where we have set $M=2$, and $N=2$. For $W<W_{c}$ we are in the diffusive side and for $W>W_{c}$ we have localization. Within this approach, $F_{n}$, $b_{m}$, $\nu$ and $W_{c}$ are free parameters that we determine using a least-square fit of the gathered data for $\Lambda$ at sufficiently long times.

\begin{figure}[t]
    \includegraphics[width=82mm]{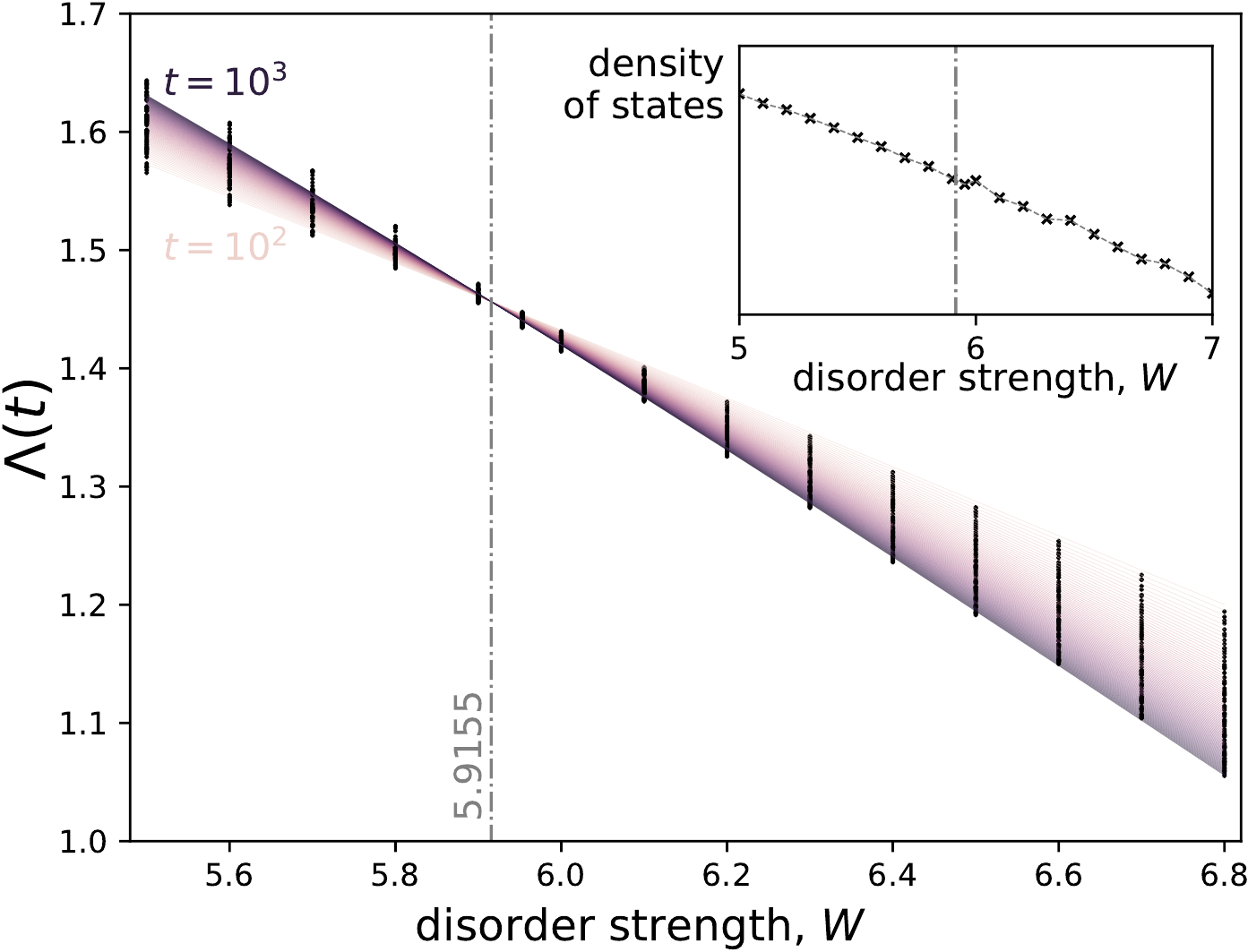}
    \caption{\label{fig:Lambda_vs_W}
            Inverse scaled CBS width $\Lambda(t)$ for times ranging from $t=10^2$ to $t=10^3$ (in units of $\hbar/J$), as functions of the disorder strength $W$ (in units of $J$). The longer times correspond to darker curves. The energy is fixed at $E = 1$ (in units of $J$). All curves cross at the mobility edge $W_c = 5.92$ and the critical exponent is $\nu = 2.74$, values that have been extracted from fitting the Taylor expansion of $F(z)$ to the numerical data (see text). The inset shows the smooth behavior of the disorder-average DoS per unit surface $\rho(E,W)$ 
            across the transition at energy $E=1$.
            }
\end{figure}

\begin{figure}[t]
    \includegraphics[width=82mm]{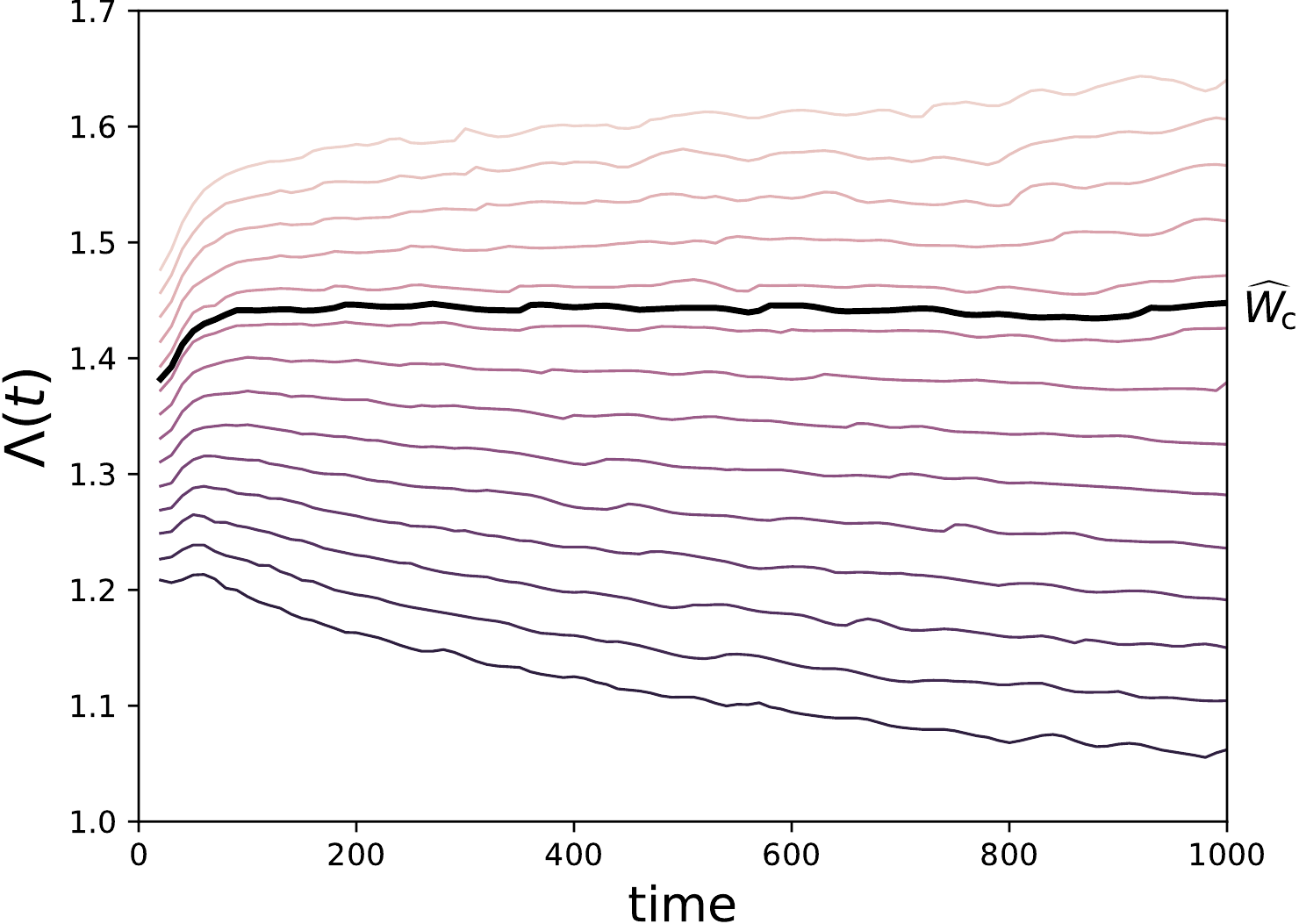}
    \caption{\label{fig:ScaledCBS_vs_Time}
            Solid lines: Inverse rescaled CBS width $\Lambda (t)$ as a function of $t$ for different values of the disorder strength $W$ at fixed energy $E=1$ (in units of $J$). Dashed lines:  fits obtained using the Taylor expansion of the one-parameter scaling function $F(z)$, plotted as a function of $t$, see text. The thick solid black line corresponds to the critical point $W=W_c=5.9155$. The critical exponent is $\nu = 2.7363$. At long times (not shown in the figure), for $W <W_c$, $\Lambda(t)$ displays plateaus, whereas for $W > W_{c}$, $\Lambda(t)$ behaves like $1/\sqrt{t}$.
            }
\end{figure}

\begin{figure}[b!]
    \includegraphics[width=82mm]{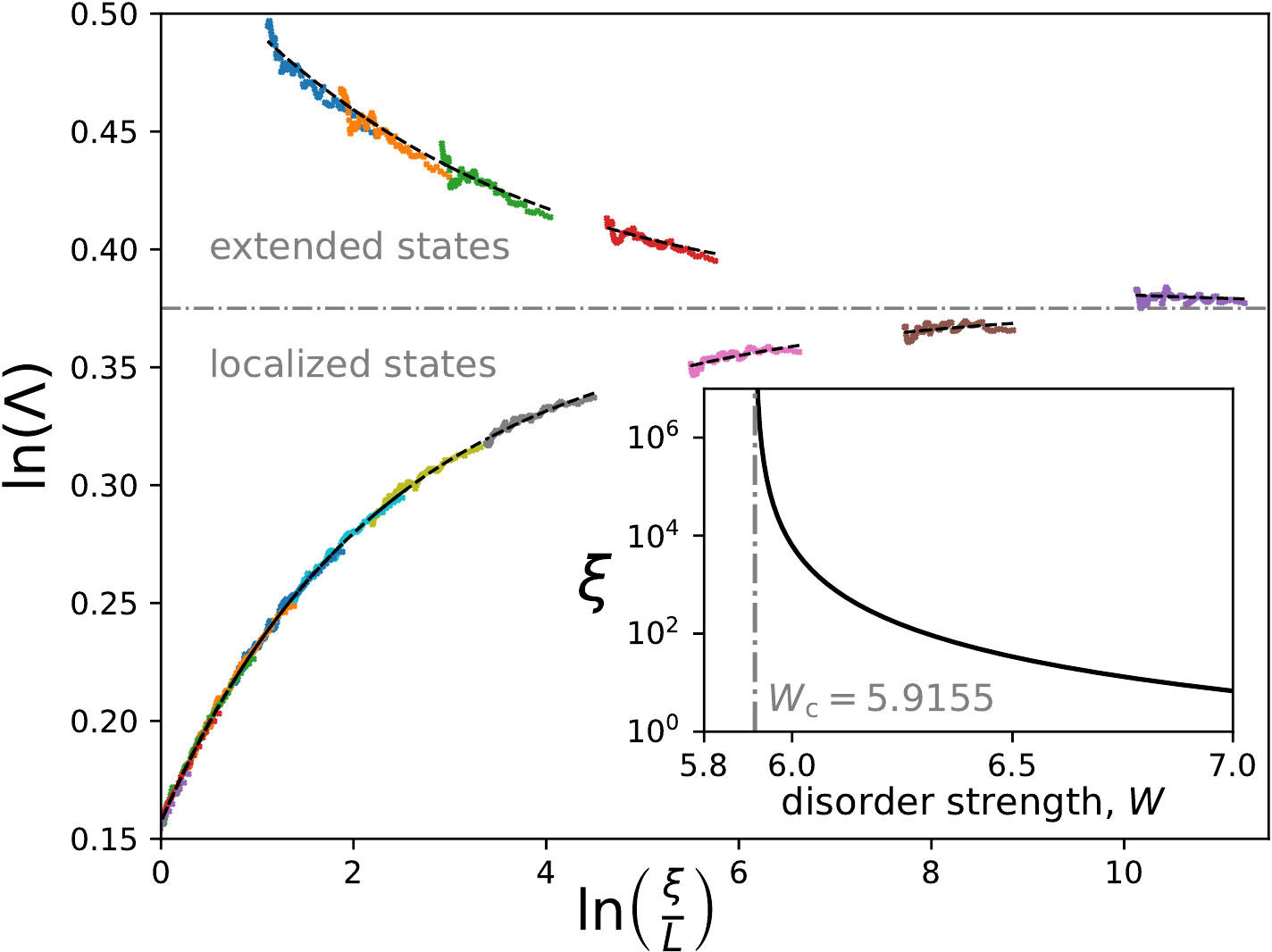}
    \caption{\label{fig:scaling_function_pub}
            Scaling function $\ln\Lambda$ as a function of $\ln(\xi/L_t)$ at energy $E=1$ (in units of $J$). The different colored pieces on the scaling curve correspond to the data obtained at different $W$ (in units of $J$). The dashed lines are the fitted curves based on the one-parameter scaling hypothesis, see text. The horizontal gray dash-dotted line marks the separation between the extended and localized branches of the scaling function. The inset shows the correlation length $\xi$ calculated from the CBS width $\Delta k$ using the fitted parameters $W_c = 5.9155$ and $\nu = 2.7363$, see text.  
            }
\end{figure}

We plot in Fig.~\ref{fig:Lambda_vs_W}, the numerical points (dots) and the fitted curves (coloured lines) from which we obtain the estimates $W_c = 5.92$ and $\nu = 2.74$ . 
In Fig.~\ref{fig:ScaledCBS_vs_Time}, we plot $\Lambda(t)$ against $t$ for different disorder strengths $W$. As one can see, for $W$ close to $W_c$, $\Lambda(t)$ is essentially constant at large enough times, showing that the CBS width $\Delta k(t)$ at the transition has the same time dependence as in Eq.\eqref{eq:CBSDif}. Note that we have actually computed the evolution at longer times, where the plateaus are much better marked. This is an interesting result because it shows that the system still exhibits a residual diffusive motion at the critical point. This observation is consistent with Wegner's law, $s = (d-2) \nu$ \cite{Wegner1976} which implies a vanishing critical exponent $s=0$ for $D \sim (W_c-W)^{s}$ in two dimensions and thus a constant diffusion coefficient. This behaviour has also been observed in \cite{PhysRevB.53.6975}.
To verify the validity of the one-parameter scaling hypothesis in this system, we have numerically extracted $\xi(E,W) = \abs{\eta(E,W)}^{-\nu}$ to collapse all data for $\Lambda (t)$, obtained at different $W$ and times, in Figs.~\ref{fig:Lambda_vs_W} and~\ref{fig:ScaledCBS_vs_Time}, on a single scaling curve~\cite{Asada2002a}. 
To construct the scaling function, in Fig.~\ref{fig:scaling_function_pub} $\ln\Lambda(t)$ is plotted as a function of $\ln (1/L_{t})$ for different disorder strength $W$ and then shifted horizontally by some quantity $\ln\xi(E,W)$ to construct a smooth continuous curves when $\ln\Lambda(t)$ is plotted as a function of $\ln (\xi/L_{t})$. The correlation length $\xi$, central to the one-parameter scaling hypothesis, identifies with the localization length $\xi_{\text{loc}}$ in the insulating phase.

\emph{Spin localization}---
Finally (not shown here), we have observed a spin localization phenomenon in the deep localized regime that we will address in a future work. In this regime, the CBS and CFS peaks become very wide. By broadening, the tails of the CBS dip decrease the background in the spin-flipping channel while the tails of the CFS peak do the opposite in the spin-preserving channel leading to an imbalanced spin population in the spin channels. Thus the system tends to retain its initial spin state in the deeply localised regime.

\emph{Conclusion.}---
We have analysed Anderson localization in an archetypical symplectic system, which is realized in a physical system if spin-orbit coupling is relevant.

We have extracted the critical exponent and the critical disorder strength using a finite-time scaling analysis of the coherent back-scattering anti-peak. The choice of a Gaussian symplectic ensemble confirms the universality of the critical exponent in the symplectic symmetry class. Such an analysis of this momentum-space signature of the phase transition is also accessible in experiment through time-of-flight expansion and absorption imaging.  
    
Furthermore, we have demonstrated that, because the CBS width scales as $t^{-1/2}$ at the transition, there exists a residual diffusion in the transition region in contrast to three-dimensional systems with a metal-insulator transition. This residual diffusion is a characteristics of any two-dimensional system in which a metal-insulator transition is observed.
    
Future work will study the Anderson transition by monitoring the CFS contrast. However, early computations have shown an additional difficulty in the localized phase: both CBS and CFS peaks exhibits slowly decaying tails in momentum space, leading to an imbalance between the two backgrounds and making an accurate measurement of the CFS contrast troublesome. We are investigating whether this imbalance is solely due to these tails or if it could be a signature of Anderson localization in the spin degrees of freedom.


\begin{acknowledgments}
    The Authors acknowledge discussions with Drs John Helm and Jean Decamp useful to the establishment of this project.
\end{acknowledgments}


\begin{thebibliography}{50}%
\makeatletter
\providecommand \@ifxundefined [1]{%
 \@ifx{#1\undefined}
}%
\providecommand \@ifnum [1]{%
 \ifnum #1\expandafter \@firstoftwo
 \else \expandafter \@secondoftwo
 \fi
}%
\providecommand \@ifx [1]{%
 \ifx #1\expandafter \@firstoftwo
 \else \expandafter \@secondoftwo
 \fi
}%
\providecommand \natexlab [1]{#1}%
\providecommand \enquote  [1]{``#1''}%
\providecommand \bibnamefont  [1]{#1}%
\providecommand \bibfnamefont [1]{#1}%
\providecommand \citenamefont [1]{#1}%
\providecommand \href@noop [0]{\@secondoftwo}%
\providecommand \href [0]{\begingroup \@sanitize@url \@href}%
\providecommand \@href[1]{\@@startlink{#1}\@@href}%
\providecommand \@@href[1]{\endgroup#1\@@endlink}%
\providecommand \@sanitize@url [0]{\catcode `\\12\catcode `\$12\catcode
  `\&12\catcode `\#12\catcode `\^12\catcode `\_12\catcode `\%12\relax}%
\providecommand \@@startlink[1]{}%
\providecommand \@@endlink[0]{}%
\providecommand \url  [0]{\begingroup\@sanitize@url \@url }%
\providecommand \@url [1]{\endgroup\@href {#1}{\urlprefix }}%
\providecommand \urlprefix  [0]{URL }%
\providecommand \Eprint [0]{\href }%
\providecommand \doibase [0]{http://dx.doi.org/}%
\providecommand \selectlanguage [0]{\@gobble}%
\providecommand \bibinfo  [0]{\@secondoftwo}%
\providecommand \bibfield  [0]{\@secondoftwo}%
\providecommand \translation [1]{[#1]}%
\providecommand \BibitemOpen [0]{}%
\providecommand \bibitemStop [0]{}%
\providecommand \bibitemNoStop [0]{.\EOS\space}%
\providecommand \EOS [0]{\spacefactor3000\relax}%
\providecommand \BibitemShut  [1]{\csname bibitem#1\endcsname}%
\let\auto@bib@innerbib\@empty
\bibitem [{\citenamefont {Anderson}(1958)}]{Anderson1958}%
  \BibitemOpen
  \bibfield  {author} {\bibinfo {author} {\bibfnamefont {P.~W.}\ \bibnamefont
  {Anderson}},\ }\href {\doibase 10.1103/PhysRev.109.1492} {\bibfield
  {journal} {\bibinfo  {journal} {Phys. Rev.}\ }\textbf {\bibinfo {volume}
  {109}},\ \bibinfo {pages} {1492} (\bibinfo {year} {1958})}\BibitemShut
  {NoStop}%
\bibitem [{\citenamefont {Slevin}\ and\ \citenamefont
  {Ohtsuki}(2014)}]{Slevin2014a}%
  \BibitemOpen
  \bibfield  {author} {\bibinfo {author} {\bibfnamefont {K.}~\bibnamefont
  {Slevin}}\ and\ \bibinfo {author} {\bibfnamefont {T.}~\bibnamefont
  {Ohtsuki}},\ }\href {\doibase 10.1088/1367-2630/16/1/015012} {\bibfield
  {journal} {\bibinfo  {journal} {New Journal of Physics}\ }\textbf {\bibinfo
  {volume} {16}},\ \bibinfo {pages} {015012} (\bibinfo {year}
  {2014})}\BibitemShut {NoStop}%
\bibitem [{\citenamefont {Ueoka}\ and\ \citenamefont
  {Slevin}(2017)}]{Ueoka2017}%
  \BibitemOpen
  \bibfield  {author} {\bibinfo {author} {\bibfnamefont {Y.}~\bibnamefont
  {Ueoka}}\ and\ \bibinfo {author} {\bibfnamefont {K.}~\bibnamefont {Slevin}},\
  }\href {\doibase 10.7566/JPSJ.86.094707} {\bibfield  {journal} {\bibinfo
  {journal} {Journal of the Physical Society of Japan}\ }\textbf {\bibinfo
  {volume} {86}},\ \bibinfo {pages} {094707} (\bibinfo {year}
  {2017})}\BibitemShut {NoStop}%
\bibitem [{\citenamefont {Evers}\ and\ \citenamefont
  {Mirlin}(2008)}]{RevModPhys.80.1355}%
  \BibitemOpen
  \bibfield  {author} {\bibinfo {author} {\bibfnamefont {F.}~\bibnamefont
  {Evers}}\ and\ \bibinfo {author} {\bibfnamefont {A.~D.}\ \bibnamefont
  {Mirlin}},\ }\href {\doibase 10.1103/RevModPhys.80.1355} {\bibfield
  {journal} {\bibinfo  {journal} {Rev. Mod. Phys.}\ }\textbf {\bibinfo {volume}
  {80}},\ \bibinfo {pages} {1355} (\bibinfo {year} {2008})}\BibitemShut
  {NoStop}%
\bibitem [{\citenamefont {Cutler}\ and\ \citenamefont {Mott}(1969)}]{Mott1969}%
  \BibitemOpen
  \bibfield  {author} {\bibinfo {author} {\bibfnamefont {M.}~\bibnamefont
  {Cutler}}\ and\ \bibinfo {author} {\bibfnamefont {N.~F.}\ \bibnamefont
  {Mott}},\ }\href {\doibase 10.1103/PhysRev.181.1336} {\bibfield  {journal}
  {\bibinfo  {journal} {Phys. Rev.}\ }\textbf {\bibinfo {volume} {181}},\
  \bibinfo {pages} {1336} (\bibinfo {year} {1969})}\BibitemShut {NoStop}%
\bibitem [{\citenamefont {Lee}\ and\ \citenamefont
  {Ramakrishnan}(1985)}]{Lee1985}%
  \BibitemOpen
  \bibfield  {author} {\bibinfo {author} {\bibfnamefont {P.~A.}\ \bibnamefont
  {Lee}}\ and\ \bibinfo {author} {\bibfnamefont {T.~V.}\ \bibnamefont
  {Ramakrishnan}},\ }\href {\doibase 10.1103/RevModPhys.57.287} {\bibfield
  {journal} {\bibinfo  {journal} {Rev. Mod. Phys.}\ }\textbf {\bibinfo {volume}
  {57}},\ \bibinfo {pages} {287} (\bibinfo {year} {1985})}\BibitemShut
  {NoStop}%
\bibitem [{\citenamefont {Weaver}(1990)}]{Weaver1990}%
  \BibitemOpen
  \bibfield  {author} {\bibinfo {author} {\bibfnamefont {R.~L.}\ \bibnamefont
  {Weaver}},\ }\href {\doibase 10.1016/0165-2125(90)90034-2} {\bibfield
  {journal} {\bibinfo  {journal} {Wave Motion}\ }\textbf {\bibinfo {volume}
  {12}},\ \bibinfo {pages} {129} (\bibinfo {year} {1990})}\BibitemShut
  {NoStop}%
\bibitem [{\citenamefont {Dalichaouch}\ \emph {et~al.}(1991)\citenamefont
  {Dalichaouch}, \citenamefont {Armstrong}, \citenamefont {Schultz},
  \citenamefont {Platzman},\ and\ \citenamefont {McCall}}]{McCall1991}%
  \BibitemOpen
  \bibfield  {author} {\bibinfo {author} {\bibfnamefont {R.}~\bibnamefont
  {Dalichaouch}}, \bibinfo {author} {\bibfnamefont {J.~P.}\ \bibnamefont
  {Armstrong}}, \bibinfo {author} {\bibfnamefont {S.}~\bibnamefont {Schultz}},
  \bibinfo {author} {\bibfnamefont {P.~M.}\ \bibnamefont {Platzman}}, \ and\
  \bibinfo {author} {\bibfnamefont {S.~L.}\ \bibnamefont {McCall}},\ }\href
  {\doibase 10.1038/354053a0} {\bibfield  {journal} {\bibinfo  {journal}
  {Nature}\ }\textbf {\bibinfo {volume} {354}},\ \bibinfo {pages} {53}
  (\bibinfo {year} {1991})}\BibitemShut {NoStop}%
\bibitem [{\citenamefont {Wiersma}\ \emph {et~al.}(1997)\citenamefont
  {Wiersma}, \citenamefont {Bartolini}, \citenamefont {Lagendijk},\ and\
  \citenamefont {Righini}}]{Wiersma1997}%
  \BibitemOpen
  \bibfield  {author} {\bibinfo {author} {\bibfnamefont {D.~S.}\ \bibnamefont
  {Wiersma}}, \bibinfo {author} {\bibfnamefont {P.}~\bibnamefont {Bartolini}},
  \bibinfo {author} {\bibfnamefont {A.}~\bibnamefont {Lagendijk}}, \ and\
  \bibinfo {author} {\bibfnamefont {R.}~\bibnamefont {Righini}},\ }\href
  {\doibase 10.1038/37757} {\bibfield  {journal} {\bibinfo  {journal} {Nature}\
  }\textbf {\bibinfo {volume} {390}},\ \bibinfo {pages} {671} (\bibinfo {year}
  {1997})}\BibitemShut {NoStop}%
\bibitem [{\citenamefont {Stoytchev}\ and\ \citenamefont
  {Genack}(1997)}]{Genack1997}%
  \BibitemOpen
  \bibfield  {author} {\bibinfo {author} {\bibfnamefont {M.}~\bibnamefont
  {Stoytchev}}\ and\ \bibinfo {author} {\bibfnamefont {A.~Z.}\ \bibnamefont
  {Genack}},\ }\href {\doibase 10.1103/PhysRevB.55.R8617} {\bibfield  {journal}
  {\bibinfo  {journal} {Phys. Rev. B}\ }\textbf {\bibinfo {volume} {55}},\
  \bibinfo {pages} {R8617} (\bibinfo {year} {1997})}\BibitemShut {NoStop}%
\bibitem [{\citenamefont {Dembowski}\ \emph {et~al.}(1999)\citenamefont
  {Dembowski}, \citenamefont {Gr\"af}, \citenamefont {Hofferbert},
  \citenamefont {Rehfeld}, \citenamefont {Richter},\ and\ \citenamefont
  {Weiland}}]{Weiland1999}%
  \BibitemOpen
  \bibfield  {author} {\bibinfo {author} {\bibfnamefont {C.}~\bibnamefont
  {Dembowski}}, \bibinfo {author} {\bibfnamefont {H.-D.}\ \bibnamefont
  {Gr\"af}}, \bibinfo {author} {\bibfnamefont {R.}~\bibnamefont {Hofferbert}},
  \bibinfo {author} {\bibfnamefont {H.}~\bibnamefont {Rehfeld}}, \bibinfo
  {author} {\bibfnamefont {A.}~\bibnamefont {Richter}}, \ and\ \bibinfo
  {author} {\bibfnamefont {T.}~\bibnamefont {Weiland}},\ }\href {\doibase
  10.1103/PhysRevE.60.3942} {\bibfield  {journal} {\bibinfo  {journal} {Phys.
  Rev. E}\ }\textbf {\bibinfo {volume} {60}},\ \bibinfo {pages} {3942}
  (\bibinfo {year} {1999})}\BibitemShut {NoStop}%
\bibitem [{\citenamefont {St\"orzer}\ \emph {et~al.}(2006)\citenamefont
  {St\"orzer}, \citenamefont {Gross}, \citenamefont {Aegerter},\ and\
  \citenamefont {Maret}}]{Storzer2006}%
  \BibitemOpen
  \bibfield  {author} {\bibinfo {author} {\bibfnamefont {M.}~\bibnamefont
  {St\"orzer}}, \bibinfo {author} {\bibfnamefont {P.}~\bibnamefont {Gross}},
  \bibinfo {author} {\bibfnamefont {C.~M.}\ \bibnamefont {Aegerter}}, \ and\
  \bibinfo {author} {\bibfnamefont {G.}~\bibnamefont {Maret}},\ }\href
  {\doibase 10.1103/PhysRevLett.96.063904} {\bibfield  {journal} {\bibinfo
  {journal} {Phys. Rev. Lett.}\ }\textbf {\bibinfo {volume} {96}},\ \bibinfo
  {pages} {063904} (\bibinfo {year} {2006})}\BibitemShut {NoStop}%
\bibitem [{\citenamefont {Topolancik}\ \emph {et~al.}(2007)\citenamefont
  {Topolancik}, \citenamefont {Ilic},\ and\ \citenamefont
  {Vollmer}}]{Topolancik2007}%
  \BibitemOpen
  \bibfield  {author} {\bibinfo {author} {\bibfnamefont {J.}~\bibnamefont
  {Topolancik}}, \bibinfo {author} {\bibfnamefont {B.}~\bibnamefont {Ilic}}, \
  and\ \bibinfo {author} {\bibfnamefont {F.}~\bibnamefont {Vollmer}},\ }\href
  {\doibase 10.1103/PhysRevLett.99.253901} {\bibfield  {journal} {\bibinfo
  {journal} {Phys. Rev. Lett.}\ }\textbf {\bibinfo {volume} {99}},\ \bibinfo
  {pages} {253901} (\bibinfo {year} {2007})}\BibitemShut {NoStop}%
\bibitem [{\citenamefont {Schwartz}\ \emph {et~al.}(2007)\citenamefont
  {Schwartz}, \citenamefont {Bartal}, \citenamefont {Fishman},\ and\
  \citenamefont {Segev}}]{Schwartz2007}%
  \BibitemOpen
  \bibfield  {author} {\bibinfo {author} {\bibfnamefont {T.}~\bibnamefont
  {Schwartz}}, \bibinfo {author} {\bibfnamefont {G.}~\bibnamefont {Bartal}},
  \bibinfo {author} {\bibfnamefont {S.}~\bibnamefont {Fishman}}, \ and\
  \bibinfo {author} {\bibfnamefont {M.}~\bibnamefont {Segev}},\ }\href
  {\doibase 10.1038/nature05623} {\bibfield  {journal} {\bibinfo  {journal}
  {Nature}\ }\textbf {\bibinfo {volume} {446}},\ \bibinfo {pages} {52}
  (\bibinfo {year} {2007})}\BibitemShut {NoStop}%
\bibitem [{\citenamefont {Hu}\ \emph {et~al.}(2008)\citenamefont {Hu},
  \citenamefont {Strybulevych}, \citenamefont {Page}, \citenamefont
  {Skipetrov},\ and\ \citenamefont {van Tiggelen}}]{Hu2008}%
  \BibitemOpen
  \bibfield  {author} {\bibinfo {author} {\bibfnamefont {H.}~\bibnamefont
  {Hu}}, \bibinfo {author} {\bibfnamefont {A.}~\bibnamefont {Strybulevych}},
  \bibinfo {author} {\bibfnamefont {J.~H.}\ \bibnamefont {Page}}, \bibinfo
  {author} {\bibfnamefont {S.~E.}\ \bibnamefont {Skipetrov}}, \ and\ \bibinfo
  {author} {\bibfnamefont {B.~A.}\ \bibnamefont {van Tiggelen}},\ }\href
  {\doibase 10.1038/nphys1101} {\bibfield  {journal} {\bibinfo  {journal}
  {Nature Physics}\ }\textbf {\bibinfo {volume} {4}},\ \bibinfo {pages} {945}
  (\bibinfo {year} {2008})}\BibitemShut {NoStop}%
\bibitem [{\citenamefont {Chab\'e}\ \emph
  {et~al.}(2008{\natexlab{a}})\citenamefont {Chab\'e}, \citenamefont
  {Lemari\'e}, \citenamefont {Gr\'emaud}, \citenamefont {Delande},
  \citenamefont {Szriftgiser},\ and\ \citenamefont {Garreau}}]{Chabe2008}%
  \BibitemOpen
  \bibfield  {author} {\bibinfo {author} {\bibfnamefont {J.}~\bibnamefont
  {Chab\'e}}, \bibinfo {author} {\bibfnamefont {G.}~\bibnamefont {Lemari\'e}},
  \bibinfo {author} {\bibfnamefont {B.}~\bibnamefont {Gr\'emaud}}, \bibinfo
  {author} {\bibfnamefont {D.}~\bibnamefont {Delande}}, \bibinfo {author}
  {\bibfnamefont {P.}~\bibnamefont {Szriftgiser}}, \ and\ \bibinfo {author}
  {\bibfnamefont {J.~C.}\ \bibnamefont {Garreau}},\ }\href {\doibase
  10.1103/PhysRevLett.101.255702} {\bibfield  {journal} {\bibinfo  {journal}
  {Phys. Rev. Lett.}\ }\textbf {\bibinfo {volume} {101}},\ \bibinfo {pages}
  {255702} (\bibinfo {year} {2008}{\natexlab{a}})}\BibitemShut {NoStop}%
\bibitem [{\citenamefont {Riboli}\ \emph {et~al.}(2011)\citenamefont {Riboli},
  \citenamefont {Barthelemy}, \citenamefont {Vignolini}, \citenamefont
  {Intonti}, \citenamefont {Rossi}, \citenamefont {Combrie},\ and\
  \citenamefont {Wiersma}}]{Riboli2011}%
  \BibitemOpen
  \bibfield  {author} {\bibinfo {author} {\bibfnamefont {F.}~\bibnamefont
  {Riboli}}, \bibinfo {author} {\bibfnamefont {P.}~\bibnamefont {Barthelemy}},
  \bibinfo {author} {\bibfnamefont {S.}~\bibnamefont {Vignolini}}, \bibinfo
  {author} {\bibfnamefont {F.}~\bibnamefont {Intonti}}, \bibinfo {author}
  {\bibfnamefont {A.~D.}\ \bibnamefont {Rossi}}, \bibinfo {author}
  {\bibfnamefont {S.}~\bibnamefont {Combrie}}, \ and\ \bibinfo {author}
  {\bibfnamefont {D.~S.}\ \bibnamefont {Wiersma}},\ }\href {\doibase
  10.1364/OL.36.000127} {\bibfield  {journal} {\bibinfo  {journal} {Opt.
  Lett.}\ }\textbf {\bibinfo {volume} {36}},\ \bibinfo {pages} {127} (\bibinfo
  {year} {2011})}\BibitemShut {NoStop}%
\bibitem [{\citenamefont {Sperling}\ \emph {et~al.}(2012)\citenamefont
  {Sperling}, \citenamefont {B\"{u}hrer}, \citenamefont {Aegerter},\ and\
  \citenamefont {Maret}}]{Sperling2012}%
  \BibitemOpen
  \bibfield  {author} {\bibinfo {author} {\bibfnamefont {T.}~\bibnamefont
  {Sperling}}, \bibinfo {author} {\bibfnamefont {W.}~\bibnamefont
  {B\"{u}hrer}}, \bibinfo {author} {\bibfnamefont {C.~M.}\ \bibnamefont
  {Aegerter}}, \ and\ \bibinfo {author} {\bibfnamefont {G.}~\bibnamefont
  {Maret}},\ }\href {\doibase 10.1038/nphoton.2012.313} {\bibfield  {journal}
  {\bibinfo  {journal} {Nature Photonics}\ }\textbf {\bibinfo {volume} {7}},\
  \bibinfo {pages} {48} (\bibinfo {year} {2012})}\BibitemShut {NoStop}%
\bibitem [{\citenamefont {Lopez}\ \emph {et~al.}(2012)\citenamefont {Lopez},
  \citenamefont {Cl\'ement}, \citenamefont {Szriftgiser}, \citenamefont
  {Garreau},\ and\ \citenamefont {Delande}}]{Lopez2012}%
  \BibitemOpen
  \bibfield  {author} {\bibinfo {author} {\bibfnamefont {M.}~\bibnamefont
  {Lopez}}, \bibinfo {author} {\bibfnamefont {J.-F.}\ \bibnamefont
  {Cl\'ement}}, \bibinfo {author} {\bibfnamefont {P.}~\bibnamefont
  {Szriftgiser}}, \bibinfo {author} {\bibfnamefont {J.~C.}\ \bibnamefont
  {Garreau}}, \ and\ \bibinfo {author} {\bibfnamefont {D.}~\bibnamefont
  {Delande}},\ }\href {\doibase 10.1103/PhysRevLett.108.095701} {\bibfield
  {journal} {\bibinfo  {journal} {Phys. Rev. Lett.}\ }\textbf {\bibinfo
  {volume} {108}},\ \bibinfo {pages} {095701} (\bibinfo {year}
  {2012})}\BibitemShut {NoStop}%
\bibitem [{\citenamefont {Manai}\ \emph {et~al.}(2015)\citenamefont {Manai},
  \citenamefont {Cl\'ement}, \citenamefont {Chicireanu}, \citenamefont
  {Hainaut}, \citenamefont {Garreau}, \citenamefont {Szriftgiser},\ and\
  \citenamefont {Delande}}]{Manai2015}%
  \BibitemOpen
  \bibfield  {author} {\bibinfo {author} {\bibfnamefont {I.}~\bibnamefont
  {Manai}}, \bibinfo {author} {\bibfnamefont {J.-F.}\ \bibnamefont
  {Cl\'ement}}, \bibinfo {author} {\bibfnamefont {R.}~\bibnamefont
  {Chicireanu}}, \bibinfo {author} {\bibfnamefont {C.}~\bibnamefont {Hainaut}},
  \bibinfo {author} {\bibfnamefont {J.~C.}\ \bibnamefont {Garreau}}, \bibinfo
  {author} {\bibfnamefont {P.}~\bibnamefont {Szriftgiser}}, \ and\ \bibinfo
  {author} {\bibfnamefont {D.}~\bibnamefont {Delande}},\ }\href {\doibase
  10.1103/PhysRevLett.115.240603} {\bibfield  {journal} {\bibinfo  {journal}
  {Phys. Rev. Lett.}\ }\textbf {\bibinfo {volume} {115}},\ \bibinfo {pages}
  {240603} (\bibinfo {year} {2015})}\BibitemShut {NoStop}%
\bibitem [{\citenamefont {Ying}\ \emph {et~al.}(2016)\citenamefont {Ying},
  \citenamefont {Gu}, \citenamefont {Chen}, \citenamefont {Wang}, \citenamefont
  {Jin}, \citenamefont {Zhao}, \citenamefont {Zhang},\ and\ \citenamefont
  {Chen}}]{Ying2016}%
  \BibitemOpen
  \bibfield  {author} {\bibinfo {author} {\bibfnamefont {T.}~\bibnamefont
  {Ying}}, \bibinfo {author} {\bibfnamefont {Y.}~\bibnamefont {Gu}}, \bibinfo
  {author} {\bibfnamefont {X.}~\bibnamefont {Chen}}, \bibinfo {author}
  {\bibfnamefont {X.}~\bibnamefont {Wang}}, \bibinfo {author} {\bibfnamefont
  {S.}~\bibnamefont {Jin}}, \bibinfo {author} {\bibfnamefont {L.}~\bibnamefont
  {Zhao}}, \bibinfo {author} {\bibfnamefont {W.}~\bibnamefont {Zhang}}, \ and\
  \bibinfo {author} {\bibfnamefont {X.}~\bibnamefont {Chen}},\ }\href {\doibase
  10.1126/sciadv.1501283} {\bibfield  {journal} {\bibinfo  {journal} {{S}cience
  {A}dvances}\ }\textbf {\bibinfo {volume} {2}},\ \bibinfo {pages} {e1501283}
  (\bibinfo {year} {2016})}\BibitemShut {NoStop}%
\bibitem [{\citenamefont {Jendrzejewski}\ \emph {et~al.}(2012)\citenamefont
  {Jendrzejewski}, \citenamefont {M\"uller}, \citenamefont {Richard},
  \citenamefont {Date}, \citenamefont {Plisson}, \citenamefont {Bouyer},
  \citenamefont {Aspect},\ and\ \citenamefont
  {Josse}}]{PhysRevLett.109.195302}%
  \BibitemOpen
  \bibfield  {author} {\bibinfo {author} {\bibfnamefont {F.}~\bibnamefont
  {Jendrzejewski}}, \bibinfo {author} {\bibfnamefont {K.}~\bibnamefont
  {M\"uller}}, \bibinfo {author} {\bibfnamefont {J.}~\bibnamefont {Richard}},
  \bibinfo {author} {\bibfnamefont {A.}~\bibnamefont {Date}}, \bibinfo {author}
  {\bibfnamefont {T.}~\bibnamefont {Plisson}}, \bibinfo {author} {\bibfnamefont
  {P.}~\bibnamefont {Bouyer}}, \bibinfo {author} {\bibfnamefont
  {A.}~\bibnamefont {Aspect}}, \ and\ \bibinfo {author} {\bibfnamefont
  {V.}~\bibnamefont {Josse}},\ }\href {\doibase 10.1103/PhysRevLett.109.195302}
  {\bibfield  {journal} {\bibinfo  {journal} {Phys. Rev. Lett.}\ }\textbf
  {\bibinfo {volume} {109}},\ \bibinfo {pages} {195302} (\bibinfo {year}
  {2012})}\BibitemShut {NoStop}%
\bibitem [{\citenamefont {Roati}\ \emph {et~al.}(2008)\citenamefont {Roati},
  \citenamefont {D'Errico}, \citenamefont {Fallani}, \citenamefont {Fattori},
  \citenamefont {Fort}, \citenamefont {Zaccanti}, \citenamefont {Modugno},
  \citenamefont {Modugno},\ and\ \citenamefont {Inguscio}}]{Roati2008}%
  \BibitemOpen
  \bibfield  {author} {\bibinfo {author} {\bibfnamefont {G.}~\bibnamefont
  {Roati}}, \bibinfo {author} {\bibfnamefont {C.}~\bibnamefont {D'Errico}},
  \bibinfo {author} {\bibfnamefont {L.}~\bibnamefont {Fallani}}, \bibinfo
  {author} {\bibfnamefont {M.}~\bibnamefont {Fattori}}, \bibinfo {author}
  {\bibfnamefont {C.}~\bibnamefont {Fort}}, \bibinfo {author} {\bibfnamefont
  {M.}~\bibnamefont {Zaccanti}}, \bibinfo {author} {\bibfnamefont
  {G.}~\bibnamefont {Modugno}}, \bibinfo {author} {\bibfnamefont
  {M.}~\bibnamefont {Modugno}}, \ and\ \bibinfo {author} {\bibfnamefont
  {M.}~\bibnamefont {Inguscio}},\ }\href {\doibase 10.1038/nature07071}
  {\bibfield  {journal} {\bibinfo  {journal} {Nature}\ }\textbf {\bibinfo
  {volume} {453}},\ \bibinfo {pages} {895} (\bibinfo {year}
  {2008})}\BibitemShut {NoStop}%
\bibitem [{\citenamefont {Kondov}\ \emph {et~al.}(2011)\citenamefont {Kondov},
  \citenamefont {McGehee}, \citenamefont {Zirbel},\ and\ \citenamefont
  {DeMarco}}]{doi:10.1126/science.1209019}%
  \BibitemOpen
  \bibfield  {author} {\bibinfo {author} {\bibfnamefont {S.~S.}\ \bibnamefont
  {Kondov}}, \bibinfo {author} {\bibfnamefont {W.~R.}\ \bibnamefont {McGehee}},
  \bibinfo {author} {\bibfnamefont {J.~J.}\ \bibnamefont {Zirbel}}, \ and\
  \bibinfo {author} {\bibfnamefont {B.}~\bibnamefont {DeMarco}},\ }\href
  {\doibase 10.1126/science.1209019} {\bibfield  {journal} {\bibinfo  {journal}
  {Science}\ }\textbf {\bibinfo {volume} {334}},\ \bibinfo {pages} {66}
  (\bibinfo {year} {2011})}\BibitemShut {NoStop}%
\bibitem [{\citenamefont {Lemari\'e}\ \emph {et~al.}(2010)\citenamefont
  {Lemari\'e}, \citenamefont {Lignier}, \citenamefont {Delande}, \citenamefont
  {Szriftgiser},\ and\ \citenamefont {Garreau}}]{PhysRevLett.105.090601}%
  \BibitemOpen
  \bibfield  {author} {\bibinfo {author} {\bibfnamefont {G.}~\bibnamefont
  {Lemari\'e}}, \bibinfo {author} {\bibfnamefont {H.}~\bibnamefont {Lignier}},
  \bibinfo {author} {\bibfnamefont {D.}~\bibnamefont {Delande}}, \bibinfo
  {author} {\bibfnamefont {P.}~\bibnamefont {Szriftgiser}}, \ and\ \bibinfo
  {author} {\bibfnamefont {J.~C.}\ \bibnamefont {Garreau}},\ }\href {\doibase
  10.1103/PhysRevLett.105.090601} {\bibfield  {journal} {\bibinfo  {journal}
  {Phys. Rev. Lett.}\ }\textbf {\bibinfo {volume} {105}},\ \bibinfo {pages}
  {090601} (\bibinfo {year} {2010})}\BibitemShut {NoStop}%
\bibitem [{\citenamefont {Chab\'e}\ \emph
  {et~al.}(2008{\natexlab{b}})\citenamefont {Chab\'e}, \citenamefont
  {Lemari\'e}, \citenamefont {Gr\'emaud}, \citenamefont {Delande},
  \citenamefont {Szriftgiser},\ and\ \citenamefont
  {Garreau}}]{PhysRevLett.101.255702}%
  \BibitemOpen
  \bibfield  {author} {\bibinfo {author} {\bibfnamefont {J.}~\bibnamefont
  {Chab\'e}}, \bibinfo {author} {\bibfnamefont {G.}~\bibnamefont {Lemari\'e}},
  \bibinfo {author} {\bibfnamefont {B.}~\bibnamefont {Gr\'emaud}}, \bibinfo
  {author} {\bibfnamefont {D.}~\bibnamefont {Delande}}, \bibinfo {author}
  {\bibfnamefont {P.}~\bibnamefont {Szriftgiser}}, \ and\ \bibinfo {author}
  {\bibfnamefont {J.~C.}\ \bibnamefont {Garreau}},\ }\href {\doibase
  10.1103/PhysRevLett.101.255702} {\bibfield  {journal} {\bibinfo  {journal}
  {Phys. Rev. Lett.}\ }\textbf {\bibinfo {volume} {101}},\ \bibinfo {pages}
  {255702} (\bibinfo {year} {2008}{\natexlab{b}})}\BibitemShut {NoStop}%
\bibitem [{\citenamefont {Billy}\ \emph {et~al.}(2008)\citenamefont {Billy},
  \citenamefont {Josse}, \citenamefont {Zuo}, \citenamefont {Bernard},
  \citenamefont {Hambrecht}, \citenamefont {Lugan}, \citenamefont
  {Cl{\'{e}}ment}, \citenamefont {Sanchez-Palencia}, \citenamefont {Bouyer},\
  and\ \citenamefont {Aspect}}]{Billy2008}%
  \BibitemOpen
  \bibfield  {author} {\bibinfo {author} {\bibfnamefont {J.}~\bibnamefont
  {Billy}}, \bibinfo {author} {\bibfnamefont {V.}~\bibnamefont {Josse}},
  \bibinfo {author} {\bibfnamefont {Z.}~\bibnamefont {Zuo}}, \bibinfo {author}
  {\bibfnamefont {A.}~\bibnamefont {Bernard}}, \bibinfo {author} {\bibfnamefont
  {B.}~\bibnamefont {Hambrecht}}, \bibinfo {author} {\bibfnamefont
  {P.}~\bibnamefont {Lugan}}, \bibinfo {author} {\bibfnamefont
  {D.}~\bibnamefont {Cl{\'{e}}ment}}, \bibinfo {author} {\bibfnamefont
  {L.}~\bibnamefont {Sanchez-Palencia}}, \bibinfo {author} {\bibfnamefont
  {P.}~\bibnamefont {Bouyer}}, \ and\ \bibinfo {author} {\bibfnamefont
  {A.}~\bibnamefont {Aspect}},\ }\href {\doibase 10.1038/nature07000}
  {\bibfield  {journal} {\bibinfo  {journal} {Nature}\ }\textbf {\bibinfo
  {volume} {453}},\ \bibinfo {pages} {891} (\bibinfo {year}
  {2008})}\BibitemShut {NoStop}%
\bibitem [{\citenamefont {Semeghini}\ \emph {et~al.}(2015)\citenamefont
  {Semeghini}, \citenamefont {Landini}, \citenamefont {Castilho}, \citenamefont
  {Roy}, \citenamefont {Spagnolli}, \citenamefont {Trenkwalder}, \citenamefont
  {Fattori}, \citenamefont {Inguscio},\ and\ \citenamefont
  {Modugno}}]{Semeghini2015}%
  \BibitemOpen
  \bibfield  {author} {\bibinfo {author} {\bibfnamefont {G.}~\bibnamefont
  {Semeghini}}, \bibinfo {author} {\bibfnamefont {M.}~\bibnamefont {Landini}},
  \bibinfo {author} {\bibfnamefont {P.}~\bibnamefont {Castilho}}, \bibinfo
  {author} {\bibfnamefont {S.}~\bibnamefont {Roy}}, \bibinfo {author}
  {\bibfnamefont {G.}~\bibnamefont {Spagnolli}}, \bibinfo {author}
  {\bibfnamefont {A.}~\bibnamefont {Trenkwalder}}, \bibinfo {author}
  {\bibfnamefont {M.}~\bibnamefont {Fattori}}, \bibinfo {author} {\bibfnamefont
  {M.}~\bibnamefont {Inguscio}}, \ and\ \bibinfo {author} {\bibfnamefont
  {G.}~\bibnamefont {Modugno}},\ }\href {\doibase 10.1038/nphys3339} {\bibfield
   {journal} {\bibinfo  {journal} {Nature Physics}\ }\textbf {\bibinfo {volume}
  {11}},\ \bibinfo {pages} {554} (\bibinfo {year} {2015})}\BibitemShut
  {NoStop}%
\bibitem [{\citenamefont {Müller}\ and\ \citenamefont
  {Delande}(2010)}]{Muller2010}%
  \BibitemOpen
  \bibfield  {author} {\bibinfo {author} {\bibfnamefont {C.~A.}\ \bibnamefont
  {Müller}}\ and\ \bibinfo {author} {\bibfnamefont {D.}~\bibnamefont
  {Delande}},\ }\href@noop {} {\enquote {\bibinfo {title} {Disorder and
  interference: localization phenomena},}\ } (\bibinfo {year} {2010}),\ \Eprint
  {http://arxiv.org/abs/arXiv:1005.0915} {arXiv:1005.0915} \BibitemShut
  {NoStop}%
\bibitem [{\citenamefont {Ghosh}\ \emph
  {et~al.}(2014{\natexlab{a}})\citenamefont {Ghosh}, \citenamefont {Cherroret},
  \citenamefont {Gr\'emaud}, \citenamefont {Miniatura},\ and\ \citenamefont
  {Delande}}]{Ghosh2014}%
  \BibitemOpen
  \bibfield  {author} {\bibinfo {author} {\bibfnamefont {S.}~\bibnamefont
  {Ghosh}}, \bibinfo {author} {\bibfnamefont {N.}~\bibnamefont {Cherroret}},
  \bibinfo {author} {\bibfnamefont {B.}~\bibnamefont {Gr\'emaud}}, \bibinfo
  {author} {\bibfnamefont {C.}~\bibnamefont {Miniatura}}, \ and\ \bibinfo
  {author} {\bibfnamefont {D.}~\bibnamefont {Delande}},\ }\href {\doibase
  10.1103/PhysRevA.90.063602} {\bibfield  {journal} {\bibinfo  {journal} {Phys.
  Rev. A}\ }\textbf {\bibinfo {volume} {90}},\ \bibinfo {pages} {063602}
  (\bibinfo {year} {2014}{\natexlab{a}})}\BibitemShut {NoStop}%
\bibitem [{\citenamefont {Ghosh}\ \emph
  {et~al.}(2015{\natexlab{a}})\citenamefont {Ghosh}, \citenamefont {Delande},
  \citenamefont {Miniatura},\ and\ \citenamefont {Cherroret}}]{Ghosh2015}%
  \BibitemOpen
  \bibfield  {author} {\bibinfo {author} {\bibfnamefont {S.}~\bibnamefont
  {Ghosh}}, \bibinfo {author} {\bibfnamefont {D.}~\bibnamefont {Delande}},
  \bibinfo {author} {\bibfnamefont {C.}~\bibnamefont {Miniatura}}, \ and\
  \bibinfo {author} {\bibfnamefont {N.}~\bibnamefont {Cherroret}},\ }\href
  {\doibase 10.1103/PhysRevLett.115.200602} {\bibfield  {journal} {\bibinfo
  {journal} {Phys. Rev. Lett.}\ }\textbf {\bibinfo {volume} {115}},\ \bibinfo
  {pages} {200602} (\bibinfo {year} {2015}{\natexlab{a}})}\BibitemShut
  {NoStop}%
\bibitem [{\citenamefont {Thouless}(1974)}]{Thouless1974}%
  \BibitemOpen
  \bibfield  {author} {\bibinfo {author} {\bibfnamefont {D.}~\bibnamefont
  {Thouless}},\ }\href {\doibase https://doi.org/10.1016/0370-1573(74)90029-5}
  {\bibfield  {journal} {\bibinfo  {journal} {Physics Reports}\ }\textbf
  {\bibinfo {volume} {13}},\ \bibinfo {pages} {93} (\bibinfo {year}
  {1974})}\BibitemShut {NoStop}%
\bibitem [{\citenamefont {Abrahams}\ \emph {et~al.}(1979)\citenamefont
  {Abrahams}, \citenamefont {Anderson}, \citenamefont {Licciardello},\ and\
  \citenamefont {Ramakrishnan}}]{Abrahams1979}%
  \BibitemOpen
  \bibfield  {author} {\bibinfo {author} {\bibfnamefont {E.}~\bibnamefont
  {Abrahams}}, \bibinfo {author} {\bibfnamefont {P.~W.}\ \bibnamefont
  {Anderson}}, \bibinfo {author} {\bibfnamefont {D.~C.}\ \bibnamefont
  {Licciardello}}, \ and\ \bibinfo {author} {\bibfnamefont {T.~V.}\
  \bibnamefont {Ramakrishnan}},\ }\href {\doibase 10.1103/PhysRevLett.42.673}
  {\bibfield  {journal} {\bibinfo  {journal} {Phys. Rev. Lett.}\ }\textbf
  {\bibinfo {volume} {42}},\ \bibinfo {pages} {673} (\bibinfo {year}
  {1979})}\BibitemShut {NoStop}%
\bibitem [{\citenamefont {Mott}(1987)}]{Mott1987}%
  \BibitemOpen
  \bibfield  {author} {\bibinfo {author} {\bibfnamefont {N.}~\bibnamefont
  {Mott}},\ }\href {\doibase 10.1088/0022-3719/20/21/008} {\bibfield  {journal}
  {\bibinfo  {journal} {{J}ournal of {P}hysics {C}: {S}olid {S}tate {P}hysics}\
  }\textbf {\bibinfo {volume} {20}},\ \bibinfo {pages} {3075} (\bibinfo {year}
  {1987})}\BibitemShut {NoStop}%
\bibitem [{\citenamefont {Asada}\ \emph {et~al.}(2005)\citenamefont {Asada},
  \citenamefont {Slevin},\ and\ \citenamefont {Ohtsuki}}]{asada2005anderson}%
  \BibitemOpen
  \bibfield  {author} {\bibinfo {author} {\bibfnamefont {Y.}~\bibnamefont
  {Asada}}, \bibinfo {author} {\bibfnamefont {K.}~\bibnamefont {Slevin}}, \
  and\ \bibinfo {author} {\bibfnamefont {T.}~\bibnamefont {Ohtsuki}},\
  }\href@noop {} {\bibfield  {journal} {\bibinfo  {journal} {Journal of the
  Physical Society of Japan}\ }\textbf {\bibinfo {volume} {74}},\ \bibinfo
  {pages} {238} (\bibinfo {year} {2005})}\BibitemShut {NoStop}%
\bibitem [{\citenamefont {Ghosh}\ \emph
  {et~al.}(2014{\natexlab{b}})\citenamefont {Ghosh}, \citenamefont {Cherroret},
  \citenamefont {Gr\'emaud}, \citenamefont {Miniatura},\ and\ \citenamefont
  {Delande}}]{Gosh2014a}%
  \BibitemOpen
  \bibfield  {author} {\bibinfo {author} {\bibfnamefont {S.}~\bibnamefont
  {Ghosh}}, \bibinfo {author} {\bibfnamefont {N.}~\bibnamefont {Cherroret}},
  \bibinfo {author} {\bibfnamefont {B.}~\bibnamefont {Gr\'emaud}}, \bibinfo
  {author} {\bibfnamefont {C.}~\bibnamefont {Miniatura}}, \ and\ \bibinfo
  {author} {\bibfnamefont {D.}~\bibnamefont {Delande}},\ }\href {\doibase
  10.1103/PhysRevA.90.063602} {\bibfield  {journal} {\bibinfo  {journal} {Phys.
  Rev. A}\ }\textbf {\bibinfo {volume} {90}},\ \bibinfo {pages} {063602}
  (\bibinfo {year} {2014}{\natexlab{b}})}\BibitemShut {NoStop}%
\bibitem [{\citenamefont {Ghosh}\ \emph
  {et~al.}(2015{\natexlab{b}})\citenamefont {Ghosh}, \citenamefont {Delande},
  \citenamefont {Miniatura},\ and\ \citenamefont {Cherroret}}]{Gosh2015a}%
  \BibitemOpen
  \bibfield  {author} {\bibinfo {author} {\bibfnamefont {S.}~\bibnamefont
  {Ghosh}}, \bibinfo {author} {\bibfnamefont {D.}~\bibnamefont {Delande}},
  \bibinfo {author} {\bibfnamefont {C.}~\bibnamefont {Miniatura}}, \ and\
  \bibinfo {author} {\bibfnamefont {N.}~\bibnamefont {Cherroret}},\ }\href
  {\doibase 10.1103/PhysRevLett.115.200602} {\bibfield  {journal} {\bibinfo
  {journal} {Phys. Rev. Lett.}\ }\textbf {\bibinfo {volume} {115}},\ \bibinfo
  {pages} {200602} (\bibinfo {year} {2015}{\natexlab{b}})}\BibitemShut
  {NoStop}%
\bibitem [{\citenamefont {Asada}\ \emph {et~al.}(2002)\citenamefont {Asada},
  \citenamefont {Slevin},\ and\ \citenamefont {Ohtsuki}}]{Asada2002a}%
  \BibitemOpen
  \bibfield  {author} {\bibinfo {author} {\bibfnamefont {Y.}~\bibnamefont
  {Asada}}, \bibinfo {author} {\bibfnamefont {K.}~\bibnamefont {Slevin}}, \
  and\ \bibinfo {author} {\bibfnamefont {T.}~\bibnamefont {Ohtsuki}},\ }\href
  {\doibase 10.1103/PhysRevLett.89.256601} {\bibfield  {journal} {\bibinfo
  {journal} {Phys. Rev. Lett.}\ }\textbf {\bibinfo {volume} {89}},\ \bibinfo
  {pages} {256601} (\bibinfo {year} {2002})}\BibitemShut {NoStop}%
\bibitem [{\citenamefont {Orso}(2017)}]{Orso2017}%
  \BibitemOpen
  \bibfield  {author} {\bibinfo {author} {\bibfnamefont {G.}~\bibnamefont
  {Orso}},\ }\href {\doibase 10.1103/PhysRevLett.118.105301} {\bibfield
  {journal} {\bibinfo  {journal} {Phys. Rev. Lett.}\ }\textbf {\bibinfo
  {volume} {118}},\ \bibinfo {pages} {105301} (\bibinfo {year}
  {2017})}\BibitemShut {NoStop}%
\bibitem [{Note1()}]{Note1}%
  \BibitemOpen
  \bibinfo {note} {For spin-$\protect \genfrac {}{}{}1{1}{2}$ system $T =
  -\protect \mathrm {i} \sigma _y \protect \tmspace +\thinmuskip {.1667em} K$ ,
  where $\sigma _y$ is the $y$-Pauli matrix and $K$ the complex conjugation
  operator. Therefore $T^{2} = -\protect \mathbbm {1}$}\BibitemShut {NoStop}%
\bibitem [{\citenamefont {Ghosh}(2017)}]{Sanjib2017}%
  \BibitemOpen
  \bibfield  {author} {\bibinfo {author} {\bibfnamefont {S.}~\bibnamefont
  {Ghosh}},\ }\emph {\bibinfo {title} {Momentum signatures of the Anderson
  transition}},\ \href {https://scholarbank.nus.edu.sg/handle/10635/137758}
  {Ph.D. thesis},\ \bibinfo  {school} {Centre for Quantum Technologies/National
  University of Singapore} (\bibinfo {year} {2017})\BibitemShut {NoStop}%
\bibitem [{\citenamefont {Martinez}\ \emph {et~al.}(2022)\citenamefont
  {Martinez}, \citenamefont {Lemarié}, \citenamefont {Georgeot}, \citenamefont
  {Miniatura},\ and\ \citenamefont {Giraud}}]{Martinez2022}%
  \BibitemOpen
  \bibfield  {author} {\bibinfo {author} {\bibfnamefont {M.}~\bibnamefont
  {Martinez}}, \bibinfo {author} {\bibfnamefont {G.}~\bibnamefont {Lemarié}},
  \bibinfo {author} {\bibfnamefont {B.}~\bibnamefont {Georgeot}}, \bibinfo
  {author} {\bibfnamefont {C.}~\bibnamefont {Miniatura}}, \ and\ \bibinfo
  {author} {\bibfnamefont {O.}~\bibnamefont {Giraud}},\ }\href@noop {}
  {\enquote {\bibinfo {title} {Coherent forward scattering as a robust probe of
  multifractality in critical disordered media},}\ } (\bibinfo {year} {2022}),\
  \Eprint {http://arxiv.org/abs/arXiv:2210.04796} {arXiv:2210.04796}
  \BibitemShut {NoStop}%
\bibitem [{\citenamefont {Ghosh}\ \emph {et~al.}(2017)\citenamefont {Ghosh},
  \citenamefont {Miniatura}, \citenamefont {Cherroret},\ and\ \citenamefont
  {Delande}}]{Ghosh2017}%
  \BibitemOpen
  \bibfield  {author} {\bibinfo {author} {\bibfnamefont {S.}~\bibnamefont
  {Ghosh}}, \bibinfo {author} {\bibfnamefont {C.}~\bibnamefont {Miniatura}},
  \bibinfo {author} {\bibfnamefont {N.}~\bibnamefont {Cherroret}}, \ and\
  \bibinfo {author} {\bibfnamefont {D.}~\bibnamefont {Delande}},\ }\href
  {\doibase 10.1103/PhysRevA.95.041602} {\bibfield  {journal} {\bibinfo
  {journal} {Phys. Rev. A}\ }\textbf {\bibinfo {volume} {95}},\ \bibinfo
  {pages} {041602(R)} (\bibinfo {year} {2017})}\BibitemShut {NoStop}%
\bibitem [{\citenamefont {Lee}\ \emph {et~al.}(2014)\citenamefont {Lee},
  \citenamefont {Gr\'emaud},\ and\ \citenamefont {Miniatura}}]{LKL2014}%
  \BibitemOpen
  \bibfield  {author} {\bibinfo {author} {\bibfnamefont {K.~L.}\ \bibnamefont
  {Lee}}, \bibinfo {author} {\bibfnamefont {B.}~\bibnamefont {Gr\'emaud}}, \
  and\ \bibinfo {author} {\bibfnamefont {C.}~\bibnamefont {Miniatura}},\ }\href
  {\doibase 10.1103/PhysRevA.90.043605} {\bibfield  {journal} {\bibinfo
  {journal} {Phys. Rev. A}\ }\textbf {\bibinfo {volume} {90}},\ \bibinfo
  {pages} {043605} (\bibinfo {year} {2014})}\BibitemShut {NoStop}%
\bibitem [{\citenamefont {Beltukov}\ and\ \citenamefont
  {Skipetrov}(2017)}]{Beltukov2017}%
  \BibitemOpen
  \bibfield  {author} {\bibinfo {author} {\bibfnamefont {Y.~M.}\ \bibnamefont
  {Beltukov}}\ and\ \bibinfo {author} {\bibfnamefont {S.~E.}\ \bibnamefont
  {Skipetrov}},\ }\href {\doibase 10.1103/PhysRevB.96.174209} {\bibfield
  {journal} {\bibinfo  {journal} {Phys. Rev. B}\ }\textbf {\bibinfo {volume}
  {96}},\ \bibinfo {pages} {174209} (\bibinfo {year} {2017})}\BibitemShut
  {NoStop}%
\bibitem [{\citenamefont {Lemari\'e}\ \emph {et~al.}(2009)\citenamefont
  {Lemari\'e}, \citenamefont {Chab\'e}, \citenamefont {Szriftgiser},
  \citenamefont {Garreau}, \citenamefont {Gr\'emaud},\ and\ \citenamefont
  {Delande}}]{PhysRevA.80.043626}%
  \BibitemOpen
  \bibfield  {author} {\bibinfo {author} {\bibfnamefont {G.}~\bibnamefont
  {Lemari\'e}}, \bibinfo {author} {\bibfnamefont {J.}~\bibnamefont {Chab\'e}},
  \bibinfo {author} {\bibfnamefont {P.}~\bibnamefont {Szriftgiser}}, \bibinfo
  {author} {\bibfnamefont {J.~C.}\ \bibnamefont {Garreau}}, \bibinfo {author}
  {\bibfnamefont {B.}~\bibnamefont {Gr\'emaud}}, \ and\ \bibinfo {author}
  {\bibfnamefont {D.}~\bibnamefont {Delande}},\ }\href {\doibase
  10.1103/PhysRevA.80.043626} {\bibfield  {journal} {\bibinfo  {journal} {Phys.
  Rev. A}\ }\textbf {\bibinfo {volume} {80}},\ \bibinfo {pages} {043626}
  (\bibinfo {year} {2009})}\BibitemShut {NoStop}%
\bibitem [{\citenamefont {Lemarié}\ \emph {et~al.}(2009)\citenamefont
  {Lemarié}, \citenamefont {Grémaud},\ and\ \citenamefont
  {Delande}}]{Lemari2009}%
  \BibitemOpen
  \bibfield  {author} {\bibinfo {author} {\bibfnamefont {G.}~\bibnamefont
  {Lemarié}}, \bibinfo {author} {\bibfnamefont {B.}~\bibnamefont {Grémaud}},
  \ and\ \bibinfo {author} {\bibfnamefont {D.}~\bibnamefont {Delande}},\ }\href
  {\doibase 10.1209/0295-5075/87/37007} {\bibfield  {journal} {\bibinfo
  {journal} {Europhysics Letters}\ }\textbf {\bibinfo {volume} {87}},\ \bibinfo
  {pages} {37007} (\bibinfo {year} {2009})}\BibitemShut {NoStop}%
\bibitem [{\citenamefont {Asada}\ \emph {et~al.}(2003)\citenamefont {Asada},
  \citenamefont {Slevin},\ and\ \citenamefont {Ohtsuki}}]{Asada2003a}%
  \BibitemOpen
  \bibfield  {author} {\bibinfo {author} {\bibfnamefont {Y.}~\bibnamefont
  {Asada}}, \bibinfo {author} {\bibfnamefont {K.}~\bibnamefont {Slevin}}, \
  and\ \bibinfo {author} {\bibfnamefont {T.}~\bibnamefont {Ohtsuki}},\ }\href
  {\doibase https://doi.org/10.1016/S1386-9477(02)01014-7} {\bibfield
  {journal} {\bibinfo  {journal} {Physica E: Low-dimensional Systems and
  Nanostructures}\ }\textbf {\bibinfo {volume} {18}},\ \bibinfo {pages} {274}
  (\bibinfo {year} {2003})}\BibitemShut {NoStop}%
\bibitem [{\citenamefont {Wegner}(1976)}]{Wegner1976}%
  \BibitemOpen
  \bibfield  {author} {\bibinfo {author} {\bibfnamefont {F.~J.}\ \bibnamefont
  {Wegner}},\ }\href {\doibase 10.1007/bf01315248} {\bibfield  {journal}
  {\bibinfo  {journal} {Zeitschrift f{\"u}r Physik B Condensed Matter and
  Quanta}\ }\textbf {\bibinfo {volume} {25}},\ \bibinfo {pages} {327} (\bibinfo
  {year} {1976})}\BibitemShut {NoStop}%
\bibitem [{\citenamefont {Kawarabayashi}\ and\ \citenamefont
  {Ohtsuki}(1996)}]{PhysRevB.53.6975}%
  \BibitemOpen
  \bibfield  {author} {\bibinfo {author} {\bibfnamefont {T.}~\bibnamefont
  {Kawarabayashi}}\ and\ \bibinfo {author} {\bibfnamefont {T.}~\bibnamefont
  {Ohtsuki}},\ }\href {\doibase 10.1103/PhysRevB.53.6975} {\bibfield  {journal}
  {\bibinfo  {journal} {Phys. Rev. B}\ }\textbf {\bibinfo {volume} {53}},\
  \bibinfo {pages} {6975} (\bibinfo {year} {1996})}\BibitemShut {NoStop}%
\end{thebibliography}
%

\end{document}